\documentclass[12pt]{article}

\newcommand{\be}{\begin{equation}}
\newcommand{\ee}{\end{equation}}

\newcommand{\ba}{\begin{eqnarray}}
\newcommand{\ea}{\end{eqnarray}}

\begin{document}

\hsize36truepc\vsize51truepc
\hoffset=-.4truein\voffset=-0.5truein
\setlength{\textheight}{8.5 in}

\begin{titlepage}
\begin{center}
\hfill \\

\hfill {LPTENS-04/13, SPhT-04/022}
\vskip 0.6 in
{\large {\bf{ ON A DYNAMICAL-LIKE REPLICA-SYMMETRY-BREAKING SCHEME FOR THE SPIN
GLASS} }}
\vskip .6 in

  {\bf C. De Dominicis$^{a)}$}{\it and} {\bf E. Br\'ezin$^{b)}$}
\end{center}
\vskip 5mm
\begin{center}
{$^{a)}$ Service de Physique Th\'eorique, CE Saclay,\\ 91191
Gif-sur-Yvette, France}\\
{\it{cirano}}@{\it{spht}}.{\it{saclay}}.{\it{cea}}.{\it{fr}}

{$^{b)}$ Laboratoire de Physique Th\'eorique, Ecole Normale
Sup\'erieure}\\ {24 rue Lhomond 75231, Paris Cedex 05,
France}{\footnote{
 Unit\'e Mixte de Recherche 8549 du Centre National de la
Recherche
Scientifique et de l'\'Ecole Normale Sup\'erieure.
 }}\\ {\it{brezin}}@{\it{corto}}.{\it{lpt}}.{\it{ens}}.{\it{fr}}\\

\end{center}

 \vskip 0.5 cm
\begin {center}{\bf Abstract}\end {center}
Considering the unphysical result obtained in the calculation of the
free-energy cost for twisting the boundary conditions in a spin glass, we
trace it to the negative multiplicities associated with the Parisi
replica-symmetry breaking (RSB). We point out that a distinct RSB, that keeps
positive multiplicities,  was proposed long ago, in the spirit of an
ultra-long time dynamical approach due to Sompolinsky. For an homogeneous
bulk system, both RSB schemes are known to yield identical free energies
and observables. However, using the dynamical RSB, we have recalculated
the twist free energy at the mean-field level. The free-energy cost of this
twist is, as expected, positive in that scheme, as it should.
\vskip 14.5pt
\end{titlepage}
\setlength{\baselineskip}{1.5\baselineskip}


\section{Introduction}
Lately a rather strange result was uncovered. As is well-known for an $O(N)$
ferromagnet, the breaking of the continuous symmetry and the associated
Goldstone modes, tax the forcing of a twist in the orientation of the
magnetization (between the $z=0$ and $z=L$ boundaries) by an amount of free
energy proportional to $L^{d-2}$\cite{BD1}. Consequently the lower critical
dimensions above which symmetry breaking occurs at a non-zero temperature is
exactly given by $d_c =2$. For a spin glass, the broken continuous symmetry
group
is the reparametrization  (or gauge) group. The analog of a rotation twist,
is now a (small) gauge twist between the $z=0$ and $z=L$ boundaries. In
contradistinction with the $O(N)$ case, through a long and arduous
calculation making use of Parisi's replica symmetry breaking (RSB)
\cite{Parisi}   on Parisi's truncated Hamiltonian, we obtained  a
twist free energy cost \cite {BD2} proportional to
$-L^{d-2+\eta}$ (i.e.with a negative coefficient)\cite{As} ; the exponent
$\eta$ is the usual order parameter anomalous dimension, computed there
to one loop. The implication for the lower critical dimension, namely
$d_c= 2-\eta\simeq 2.5$, was indeed  in agreement with previous estimates
\cite {Ka, Franz}. However the negative sign of the coefficient, i.e. a
gain in free energy under twist, was very puzzling. Taking this at
face value could point to the instability of
a solution with a space inhomogeneous order parameter. If one considers
for instance an Ising antiferromagnet, one may obtain a lower free energy
with twisted, i.e. antiparallel boundary conditions. But, given that
Parisis's space-homogeneous solution is now proven to give the exact free
energy
\cite{Guerra, Talagrand}, this seems unlikely.

A way out of this puzzle seems to be the following. Parisi's solution is
(semi) stable, all eigenvalues of the associated Hessian being non-negative
when the number of steps R of RSB goes to infinity \cite{DKT}. However the
multiplicities of those eigenvalues are all {\it{negative}}. This means that
the saddle-point where the free-energy is calculated has the characteristics of
a maximum ; hence small excursions away from it will yield an unphysical
negative free-energy cost.  This situation, in turn, is due to the fact that,
in Parisi's RSB, the natural ordering of the box sizes $p_u$ (where $u$ is
the discrete overlap) is {\it{reversed}}, i.e. one works with
$$ p_{R+1} \equiv  1\geq p_R\geq \cdots\geq p_u\geq \cdots \geq p_1\geq
p_0\equiv n.$$ In this article we would like to reconsider   this question
of the
free-energy cost under a twist of boundary conditions, in the light of a
distinct
RSB scheme proposed long ago\cite{DGO}. This alternative scheme has the
following
characteristics :
\begin{itemize}
\item The box order is {\it{not}} reversed, hence one has  multiplicities
remaining  positive:
$$ p_0\gg p_1\gg \cdots \gg p_u\gg \cdots  p_R\gg p_{R+1} \equiv 1$$
\item Whereas $p_u$ in Parisi's approach is a parameter without any physical
conjugate field, it is replaced here by a susceptibility derivative
$\dot\Delta_u$, a physical quantity associated with a
dynamical-like approach to the spin-glass problem (whose history is sketched
below).  In fact it is a better candidate for an order
parameter, since it vanishes above the Almeida-Touless \cite{AT} line and is
non-zero below.
\item
At the saddle-point it gives exactly the same value as Parisi's for all
observables. It yields also the same eigenvalues for the Hessian at the saddle
point \cite{KD}, but with different individual multiplicities.
\item
The cost of a twist becomes now {\it{positive}}, at least to lowest
order  ( for the "kinetic" part of the free energy, i.e. the part
which depends of the spatial variation of the order parameter, since the
potential energy is invariant under the twist). The one-loop contribution will
undoubtedly take some time to sort out.
\end{itemize}
To replace in context  the choice made here, a short historical reminder seems
appropriate. It all started with Sommers \cite{S}  proposing a new solution
with non-negative entropy to the Sherrington-Kirkpatrick \cite {SK}
model, a solution possessing a linear response anomaly.
Enlarging on the RSB proposal of Blandin et al. \cite{Blandin}, soon
after,  De Dominicis and Garel
\cite{DG}, Bray and Moore \cite {BM} linked the non-negativity of the entropy
to the infinite limit of what has been called above the box size $p_0$, a
limit that restituted \cite{BM} the Sommers solution. Substituting opaque
replicas by dynamics \cite{DD} led to some physical  insight and Sompolinsky
\cite{Sompo} proposed to describe spin glasses in terms of the spin
correlation function
$q= \langle \sigma\sigma\rangle$ and a spin susceptibility, or linear
response $\chi= \langle \hat{\sigma}\sigma\rangle$ ($\hat{\sigma}$ being
coupled to a magnetic field).
 The linear response anomaly $\Delta$ is closely related to $\chi$. In the
long time limit (when the initial time in Langevin's equations is sent to
minus infinity) he described the sytem through an infinite set of relaxation
times $\tau_u$, with $u=1,2,\cdots R$ and $R\to \infty$. When $u$ remains
discrete , his $q(\tau_u)$ is what we call $q_u$ and his $\Delta(\tau_u^{-1})$,
our $\Delta_u$ (with $-\dot\Delta_u = \Delta_u- \Delta_{u+1})$. The approach
used in the present article is thus a replica reformulation of Sompolinsky's
dynamics as in \cite{DGO}. At this point one may ask why not retain dynamics
and drop replicas altogether. The answer is that \begin{itemize}
\item although some
results have been obtained beyond mean-field through dynamics by Sompolinsky
and Zippelius \cite
{SZ,Z}, it is more difficult to work with four time variables than with four
replicas. For instance, the spectrum of masses (the eigenvalues of the Hessian)
are easily obtained with replicas, but not yet fully sorted out within the
dynamics. \item Besides, one is interested in understanding  how to obtain a
physical answer for the twist free-energy.
\end{itemize}

\noindent In the following we first compute in some details the contribution to
the free energy cost of a twist in the parametrization, using Parisi's RSB, as
already sketched before \cite{BD2}. Then we perform the same calculation for
the RSB introduced by De Dominicis, Garel and Orland \cite{DGO} (DGO), that we
might call also  "dynamical-like" for lack of a better name, getting then the
opposite sign. Finally we show that this result may also be understood from
the fact that there is a plateau contribution in Parisis's solution, where the
overlap fuction remains constant and equal to its Edwards-Anderson value, and
does not fluctuate. In the dynamical-like approach there is no such
plateau. Hence when considering spatial variations the two
approaches give distinct reults, whereas they lead to identical results
for the bulk mean-field problem.

\section {Kinetic free energy \`a la Parisi}

In terms of $n\times n$ matrices $q_{ab}$ ($q_{aa} =0$) the free energy, in
Parisi's truncated model, reads  for a spatially homogeneous order
parameter
\be nF^{(P)}/L^d = -\sum_{ab} (\frac{\tau}{2} q_{ab}^2 + \frac{g}{12} q_{ab}^4)
+ \frac{w}{6} \sum_{abc} q_{ab}q_{bc}q_{ca}  .\ee
The contribution of the kinetic part of the free energy $F_K$ (i.e. the part
which comes form a non-spatially homogeneous order parameter) is given by
\be F_K^{(P)} =  \frac{L^{d-1}}{4n} \sum_{i=0}^{L-1} \sum_{a,b} ( q_{ab}(i) -
q_{ab}(i+1))^2\ee
in which we assume twisted boundary conditions in the z-direction (and
periodic boundary conditions in the remaining (d-1) dimensions) ; space in the
z-direction is kept here discrete with  $0\leq i\leq L$.
Using Parisi's overlap function $q_u$ , $u=0,1,\cdots, R$ (with
$q_{aa}\equiv q_{R+1} = 0$), and the associated box size $p_u$ (with
$p_0\equiv n,
p_{R+1} \equiv 1$), one obtains
\be \label{3} \frac{1}{n} \sum_{ab} q_{ab}^2(i) = \sum_{u=0}^{R+1} (p_u(i)
-p_{u+1}(i)) q_u^2(i)\ee
One needs also
\be \label{4} \frac{1}{n} \sum_{ab} q_{ab}(i)q_{ab}(i+1)  =
\sum_{u=0}^{R+1}[
\frac{1}{2}(p_u(i)+p_u(i+1)) -\frac{1}{2}(p_{u+1}(i)+ p_{u+1}(i+1))]
q_u(i)q_u(i+1)
\ee
This gives then
\be F_K^{(P)} =\frac{L^{d-1}}{4} \sum_{i=0}^{L-1} \sum_{u=0}^{R+1} [\big(p_u(i)
-p_{u+1}(i)\big) q_u(i) - \big( p_u(i+1)-p_{u+1}(i+1)\big) q_u(i+1)][q_u(i)-
q_u(i+1)]
\ee

In the $R\to \infty$ continuum limit , we have
\be p_u(i)-p_{u+1}(i) \to - \dot p(u;i) du \ee
and finally
\be \label{7} F_K^{(P)} = - \frac{L^{d-1}}{4}
\int_0^Ldz\int_0^{1-\epsilon} du
\frac{\partial}{\partial z} (\dot p(u ;z) q(u ;z)) \frac{\partial}{\partial
z}  q(u ;z). \ee At the bulk saddle-point one has
\be q(u) = \frac {w}{2g} p(u) .\ee
For twisted boundary conditions, to lowest order in the twist $h(u)$ (
$h\ll 1$ and limited to a small support  $0<u<\tilde x$),
\be \label{6}q(u ;z)=  \frac {w}{2g} (u + \frac{z}{L}
h(u)) + O(h^2)  = \frac {w}{2g} p(u;z)  + O(h^2) \ee Substituting
(\ref{6}) into
(\ref{3}) we finally obtain
\be   F_K^{(P)} = - \frac{1}{8}(\frac{w}{2g})^2 L^{d-2} \int_0^{\tilde x} du \
h^2(u)\ee that is a {\it{negative}} cost in free energy for the introduction
of a twist.

\section {Kinetic free energy \`a la DGO}
Whereas the zeroth step in Parisi approach
is the ($R=0$)  matrix $q_{ab} = q $ for $a\neq b$ (with $q_{aa} = 0$), here
the zeroth step is the so-called Sommers starting point (formally identical to
the $R=1$ Parisi's RSB). Namely, one divides the $n\times n$ matrix $Q_{ab} $
into $n /p_0$ blocks $q_{\alpha,\beta}$  (of size $p_0\times p_0$) along the
diagonal, and $\frac{n}{p_0}(\frac{n}{p_0}-1)$ off-diagonal blocks $r_{\alpha,
\beta}$ (of the same size $p_0\times p_0$) .  Then one does $R$ steps of RSB on
{\it{both    matrices}} $q_{\alpha, \beta} $ and $ r_{\alpha, \beta}$
(whereas in
Parisi's approach those steps would only invlove the diagonal blocks $
q_{\alpha,\beta}$).  One thus needs a double labelling for each element of the
initial matrix $Q_{ab} $ in order to specify for $a$ (and $b$) the
$p_0\times p_0$ block matrix under consideration, and the element in this block
matrix . One thus writes $a= (\alpha ,x)$ with $x= 1,2,\cdots , n/p_0$ and
$\alpha = 1,2,
\cdots, p_0$
\be Q_{ab} \equiv Q_{\alpha,x ; \beta,y} = q_{\alpha,\beta} \delta_{x,y} +
r_{\alpha,\beta} (1-\delta_{x,y}) .\ee
The RSB steps apply now to the matrices $q_{\alpha,\beta}$ ( with
$q_{\alpha, \alpha} = 0$) generating $q_0,q_1, \cdots, q_R, q_{R+1} \equiv
0$, and to the matrices $r_{\alpha, \beta}$ giving rise to $r_0,r_1, \cdots,
r_{R+1}$. The successive steps of RSB involve   the box sizes
$p_0,p_1, \cdots, p_R$ ($p_{R+1} \equiv  1$). In the DGO scheme the box
sizes are
made to go to infinity in the prescribed natural order
\be \label {12} p_0\gg p_1\gg \cdots \gg p_R\gg p_{R+1} \equiv 1 \ee
with at the same time letting the matrix elements of $ q- r$ go
to zero in such a way that
\be - \dot \Delta_0 = p_0(q_0-r_0) \ee
remains finite
and in general
\be -\dot \Delta_u = p_u [(q_u-r_u) -(p_{u-1}-r_{u-1})] \simeq p_u(q_u-r_u)\ee
Here $\dot \Delta_u$ can be shown to be the (discrete) susceptibility
derivative
\be -\dot \Delta_u = \Delta_u-\Delta_{u+1}
\ee
($-\dot\Delta_u$ is a positive
quantity).

As a result the Parisi free energy functional $F^{(P)}\{p_u; q_u\}$  is
replaced by a dynamical-like free energy
$F^{(D)}\{\dot\Delta_u; q_u\}$ with the associated stationarity conditions
determining $q_u, \dot \Delta_u$ and their relationship.

This alternative formulation   gives {\it{in the $R\to \infty$
continuum and at
the saddle}}-{\it{point}} results that are identical to Parisi's. Besides
it also leads to the same eigenvalues of the Hessian at the saddle-point
\cite {KD}.

After this brief reminder  we now proceed to derive the kinetic
contribution :
\be F_K^{(D)}= \frac{L^{d-1}}{4n} \sum_i \sum_{a,b} [
Q_{ab}(i)-Q_{ab}(i+1)]^2.\ee We have :
\be\frac{1}{n} \sum_{ab} Q^2_{ab} = \frac{1}{n}  \sum_{\alpha, \beta}
\sum_{x,y} [(q_{\alpha, \beta}- r_{\alpha, \beta})\delta_{x,y} + r_{\alpha,
\beta}]^2 \ee
Summing upon $x,y$ one obtains
\be  \frac{1}{n} \sum_{ab} Q^2_{ab} =
 \frac{1}{n}  \sum_{\alpha, \beta}\{\frac{n}{p_0}[ (q_{\alpha,
\beta}-r_{\alpha, \beta})^2 +2r_{\alpha,
\beta}(q_{\alpha, \beta}-r_{\alpha, \beta}) ]+ (\frac{n}{p_0})^2 r^2_{\alpha,
\beta}\} \ee
After $R$ steps of RSB one obtains
\ba \label{19} \frac{1}{n} \sum_{ab} Q^2_{ab} &=&
\frac{1}{n}\frac {n}{p_0} p_0  \{
\sum_{u=0}^R (p_u-p_{u+1}) [ (q_u^2-r_u^2) + \frac{n}{p_0} r_u]\nonumber\\
&=& \sum_{u=0}^R  p_u [(q_u^2-r_u^2) -(q^2_{u-1}-r^2_{u-1})] - r^2_{R+1}
\ea
Since $p_u\to \infty$, $ \frac{p_u}{p_{u-1}} \to 0$,
$q_u-r_u\sim\frac{1}{p_u}$ for
$u=1,\cdots,R$ in the limit (\ref{12}) the equation (\ref{19}) reduces to
\be \frac{1}{n} \sum_{ab} Q_{ab}^2 = -2 \sum _{u=0}^R q_u\dot\Delta_u -
r_{R+1}^2
\ee
The analog of (\ref{3}) is now
\be \frac{1}{n} \sum_{ab} Q_{ab}^2(i) = -2 \sum _{u=0}^R q_u(i)\dot\Delta_u
(i) - r_{R+1}^2 .\ee Likewise (\ref{4}) becomes
\be \frac{1}{n} \sum_{ab} Q_{ab}(i)Q_{ab}(i+1) = \sum_{u=0} ^R p_u[q_u(i)
q_u(i+1) -
r_u(i)r_u(i+1)] -r_{R+1}^2\ee
The box sizes going to infinity do not carry a space index any more and we end
up with
\be \frac{1}{n} \sum_{ab} Q_{ab}(i)Q_{ab}(i+1) =
\sum_{u=0} ^R [\dot  \Delta_u(i) q_u(i+1)  +\dot \Delta_u(i+1) q_u(i)]
-r_{R+1}^2 \ee

 The result for the kinetic part of the mean field free energy  functional
is thus
\ba F_K^{(D)}&=& \frac { L^{d-1}}{4n}
\sum_{i=0}^{L-1}\sum_{ab} \big(Q_{ab}(i)-Q_{ab}(i+1)
\big)^2 \nonumber \\ &=& - \frac { L^{d-1}}{2} \sum_{i=0}^L\sum_{u=0} ^R [
q_u(i)\dot\Delta_u(i)+
q_u(i+1)\dot \Delta_u(i+1)- q_u(i)\dot\Delta_u(i+1)- q_u(i+1)\dot \Delta_u(i)]
\nonumber \\ &=& - \frac
{ L^{d-1}}{2} \sum_{i=0}^L\sum_{u=0} ^R [ \big(q_u(i)-q_u(i+1)\big)
\big(\dot\Delta_u(i)-\dot\Delta_u(i+1)\big)] \ea
and in the double continuum limit, in which both space is continuous and R,
the number of steps of RSB, goes to infinity
\be F_K^{(D)} = -\frac{L^{d-1}}{2} \int_0^Ldz \int_0^{x_1} du
 \frac{\partial q(u ;z)}{\partial z}\frac{\partial \dot\Delta(u
;z)}{\partial z} \ee
At the saddle-point one has \cite{DGO, Sompo}
\be - \dot\Delta(u) = \frac{2g}{w} q(u)\dot q(u) \ee
which is the "anomalous fluctuation-dissipation relationship" for the
spin-glass.
Again, to lowest order in the twist h(t), one can just keep, as in (\ref 6)
\be q(u ;z) = \frac{w}{2g} [u + \frac{z}{L} h(u)] + O( h^2) \ee
from which one obtains, to lowest order,  the free energy for that twist
\be F_K^{(D)} = \frac{1}{4}(\frac{w}{2g})^2 L^{d-2} \int_0^{\tilde x} du\
h^2(u)
\ee so that
\be \label {26} F_K^{(D)} = -2 F_K ^{(P)} \ee
The signs are opposite : the free energy cost under twist is {\it{positive}}
 for the dynamical-like RSB.
\section { A simple calculation at the saddle-point}

Given the previous  difference between the two RSB schemes  when one considers
the spatial variation of the order parameter, one may examine in the light
of the
previous calculation why the two free energies happen to coincide for a
spatially
uniform order parameter.

For instance let us consider the bulk  contribution to the free energy per unit
 volume which is quadratic in the order parameter :
\be f_{\tau} = \frac{\tau}{4} \sum_{ab} q_{ab}^2 \ee
In Parisi's scheme one finds
\ba \label {A}  f_{\tau}^{(P)}&=&\frac{\tau}{4}[ \sum_{u=0}^{R-1}
(p_u-p_{u-1})q_{u}^2  + (p_R-1)q_R^2\nonumber \\ &=&  \frac{\tau}{4}[
-\int_{0}^{x_1}  du q^2(u)  +(x_1-1) q^2(x_1) ]\nonumber \\ &= &
\frac{\tau}{4}(\frac{w}{2g})^2[ -\frac{x_1^3}{3} + (x_1-1)x_1^2 ] \ea

Here the second term in the bracket comes from the " plateau " sector of
$q(u)$
($x_1\leq u < 1$) , at which it remains fixed to its Edwards-Anderson value
\be q(x_1) = \frac{w}{2g} x_1\ee
The mean field stationnarity conditions  determine $x_1$ in terms of the
external parameters (temperature and coupling constants).

For the dynamical-like approach one finds
\ba\label {B}  f_{\tau}^{(D)}& = &\frac{\tau}{4}[ -\sum_{u=0}^{R}
 2q(u)\dot\Delta(u) -r_{R+1}^2 \nonumber \\ &=&  \frac{\tau}{4}[
2\int_{u=0}^{x_1}  du q^2(u)\frac
{2g}{w} \dot q(u)-q^2(x_1)]  \nonumber \\ &= &
\frac{\tau}{4}(\frac{w}{2g})^2[ \frac{2x_1^3}{3} -x_1^2 ]
\ea The two results (\ref A) and ({\ref B}) at the end coincide as announced,
however the origin of the various terms are different. Indeed in (\ref A)
part of
the answer comes from the plateau where fluctuations are not allowed under
twist.
Thus when considering fluctuations the plateau part in (\ref A) will not
contribute  and one will obtain
\be (f_{\tau}^{(D)})_{fluct.} =  -2 (f_{\tau}^{(P)})_{fluct.} \ee
as we found in (\ref{26}).

Similar calculations for the parts of the free energy which are cubic and
quartic in the order parameter would  show a similar conspiracy to make
the bulk contributions identical, but yield different results when the
order parameter varies in space.

\section{Conclusion}
We have thus demonstrated  that two approaches, using different RSB schemes,
that provide identical results for the bulk mean-field energy, lead to distinct
(and opposite) answers when one enforces spatial variations of the order
parameter. The negative multiplicities occuring in Parisi's scheme were
responsible for a decrease of the free energy under twist, whereas the
dynamical-like DGO scheme does lead to an increase under twist.

This has been established only at the mean-field level. It would be
conforting to extend this to one-loop calculations at least, as we have
done in a previous work \cite{BD2} for Parisi's scheme, but the algebra,
although well-defined, is quite elaborate.

Given that the two RSB schemes yield the same bulk
 mean-field free energy, but are already different under twist, it becomes of
great interest to compare the  effect of higher loops on the bulk free-energy.

 {\bf{Acknowledgments}}:
We would like to thank G. Parisi and S. Franz for useful discussions.
One of us (CDD) would like to thank also G. Biroli and A. Crisanti.

\end{document}